\documentclass[aps,prd,12pt,preprint,tightenlines,unsortedaddress,
amsfonts,amssymb,amsmath,byrevtex,showpacs,nofootinbib]{revtex4-1}

\usepackage{latexsym}
\usepackage{graphicx}
\usepackage{geometry}
\usepackage{epstopdf}
\usepackage{color}

\usepackage{xcolor}
\usepackage{pdfsync}
\usepackage{fancyhdr}
\usepackage[applemac]{inputenc}
\usepackage[breaklinks,colorlinks,backref]{hyperref}
\hypersetup{
    colorlinks, 
    linktoc=all, 
    linkcolor=red,  
  citecolor=hyptxt,
  urlcolor=blue}
  \hypersetup{
  citebordercolor=,
  filebordercolor=green,
  linkbordercolor=blue
}
\definecolor{hyptxt}{rgb}{0.7, 0.4, 0.9}
\usepackage{bibtopic}

\newcommand{\ud}{\mathrm{d}}
\newcommand{\be}{\begin{equation}}
\newcommand{\ee}{\end{equation}}

\newcommand{\R}{\mathbb R}

\newcommand{\ket}[1]{|\kern.3ex#1\kern.3ex\rangle}
\newcommand{\bra}[1]{\langle\kern.3ex #1 \kern.3ex|}
\newcommand{\scalar}[2]{\langle\kern.3ex #1 \kern.3ex|\kern.3ex#2\kern.3ex\rangle}
\newcommand{\norm}[1]{\|\kern.3ex#1\kern.3ex \|}

\def\lg{\langle }
\def\rg{\rangle }

\def\ud{\mathrm{d}}
\def\mfn{\mathfrak{n}}

\begin{document}

\title{Vibronic framework for quantum mixmaster universe}

\author{Herv\'{e} Bergeron}
\email{herve.bergeron@u-psud.fr} \affiliation{ISMO, CNRS, Universit\'{e} Paris-Sud, Universit\'{e} Paris-Saclay,
91405 Orsay, France}

\author{Ewa Czuchry}
\email{eczuchry@fuw.edu.pl} \affiliation{National Centre for Nuclear Research, Ho{\.z}a 69,
00-681 Warszawa, Poland}

\author{Jean-Pierre Gazeau}
\email{gazeau@apc.univ-paris7.fr}
\affiliation{APC, Univ Paris Diderot, CNRS/IN2P3, CEA/Irfu, Obs de Paris, Sorbonne Paris Cit\'e, France}
\affiliation{Centro Brasileiro de Pesquisas Fisicas
22290-180 - Rio de Janeiro, RJ, Brazil }

\author{Przemys{\l}aw Ma{\l}kiewicz}
\email{pmalk@fuw.edu.pl}
\affiliation{National Centre for Nuclear Research,  Ho{\.z}a 69,
00-681
Warszawa, Poland}
\affiliation{APC, Univ Paris Diderot, CNRS/IN2P3, CEA/Irfu, Obs de Paris, Sorbonne Paris Cit\'e, France}

\date{\today}


\begin{abstract}
Following our previous papers concerning the quantization of the vacuum Bianchi-IX model within or beyond the Born-Oppenheimer and adiabatic approximation, we develop a more elaborate analysis of the dynamical properties of the model based the vibronic approach utilized in molecular physics. As in the previous papers, we restrict our approach  to the harmonic approximation of the anisotropy potential in order to obtain resoluble analytical expressions.
\end{abstract}


\maketitle

\tableofcontents

\section{Introduction}
\label{intro}

Our previous results concerning the Bianchi Type IX model \cite{Bergeron2015short, Bergeron2015long} show that the singularity of the classical theory may be replaced by a non-singular dynamics due to a consistent quantization of the gravitational field. The singularity resolution is issued from the quantization respecting the symmetries of the phase space. Notably, we employed the so-called affine coherent states (ACS) to define the quantization map.

If the range of canonical variables is the full plane, the phase space symmetry is translational, represented by the Weyl-Heisenberg group, and the quantization is canonical. This is the case for anisotropic variables, which describe aspherical deformations to the spatial geometry of the Bianchi-IX model. Thus, we quantize these variables following the usual canonical prescription. However, if the range of canonical variables is the half plane, the phase space symmetry is respected by dilation and translation, which generate the $ax+b$ affine group of the real line. Covariant quantization respecting this symmetry is obtained with the use of coherent states constructed via a unitary irreducible representation of the affine group. This is the case for isotropic variables, representing the volume and the mean expansion in the Bianchi-IX model. We quantize these variables with ACS. As a result, a quantum term, which regularizes the dynamics near the boundary of the phase space, is issued. This term is responsible for resolving the singularity of the classical dynamics and replacing it by a bounce.

To solve the quantum dynamics for the mixmaster universe, we apply assumptions inspired by approaches in molecular physics \cite{yakorny12}. In our earlier papers \cite{Bergeron2015short, Bergeron2015long} we assumed the adiabatic dynamics and employed the Born-Huang-Oppenheimer approximation. The anisotropic oscillations were approximated by a fixed quantum state during the evolution. In our recent paper \cite{B9vibronica} we made a step beyond the adiabatic approximation and allowed for excitations and decays of the oscillations. The framework included however the assumption of no backreaction from the transitions between oscillatory states on the isotropic expansion. We were able to define and solve a unitary dynamics of the model within the harmonic approximation to the anisotropy potential. A parameter describing stiffness of the bounce in the isotropic variables was recognized crucial to the dynamical properties of the anisotropic oscillations. The adiabatic approximation was found to break down when this parameter exceeds its critical value. Then a non-linear excitation of anisotropic eigenstates takes place throughout the bounce. The application of this result to a model of radiation-filled universe indicates a possibility of large production of anisotropy at the bounce  which led to some sort of a sustained phase of accelerated expansion resembling the standard inflationary phase. This quasi-inflationary phase came as a result of our quantization method based on the ADM canonical formalism and ordinary quantum mechanics, in contrast with inflationary models where an additional scalar field in a suitable potential is introduced.

In the present paper we develop a suitable framework to address the problem of the bounce followed by the quasi-inflationary phase in detail. It is called the vibronic framework and it accommodates the full evolution of the oscillatory degrees of freedom and their backreaction on the background dynamics. The background dynamics is given a semiclassical treatment by confining it to the space of coherent states. In the limit of large volumes, the semiclassical dynamics coincides with the classical one. Like in our previous papers, we apply the harmonic approximation to the anisotropy potential. We arrive at a consistent set of equations, which include quantum and semiclassical degrees of freedom and which preserve the semiclassical constraint. We illustrate our framework with a numerical  study of two examples of initial conditions. They confirm the possibility of stable Friedmann-like adiabatic quantum dynamics as well as of the breakdown of adiabatic behavior. The application of the framework to physically plausible settings requires much more involved numerical computations and is postponed to future work.

The structure of the paper is as follows. Section \ref{b9model} is a reminder of the classical and quantum Hamiltonian for the mixmaster universe. Section \ref{appsclag} extends Klauder's approach to set up the semiclassical Lagrangian framework for constrained systems with quantum internal degrees of freedom. Section \ref{discframe} is devoted to a discussion of various possible approximations for Hamiltonians with internal degrees of freedom among which the Born-Oppenheimer approximation is the simplest one and the vibronic approach is the most elaborate one. Section \ref{dyneq} concerns the specification of the vibronic approximation to the mixmaster universe within the harmonic approximation. We derive a complete set of equations of motion and we discuss the modified Friedmann equation.  In Section \ref{comdyneq} we discuss various aspects of the employed formalism and derived from it equations of motion. We integrate numerically the derived equation of motion in Section \ref{numsim} where we confirm the stability of the adiabatic approximation to the solution with anisotropy in the ground state. We also find a weakening of the stability as the anisotropy eigennumber increases, which is with agreement with our result in \cite{B9vibronica}. We conclude in Section \ref{conclu} where we discuss some physical implications of our present result and pose interesting questions to be examined in future papers.

\section{Mixmaster universe}
\label{b9model}

\subsection{Classical model}
\begin{figure}[!ht]
\includegraphics[scale=0.8]{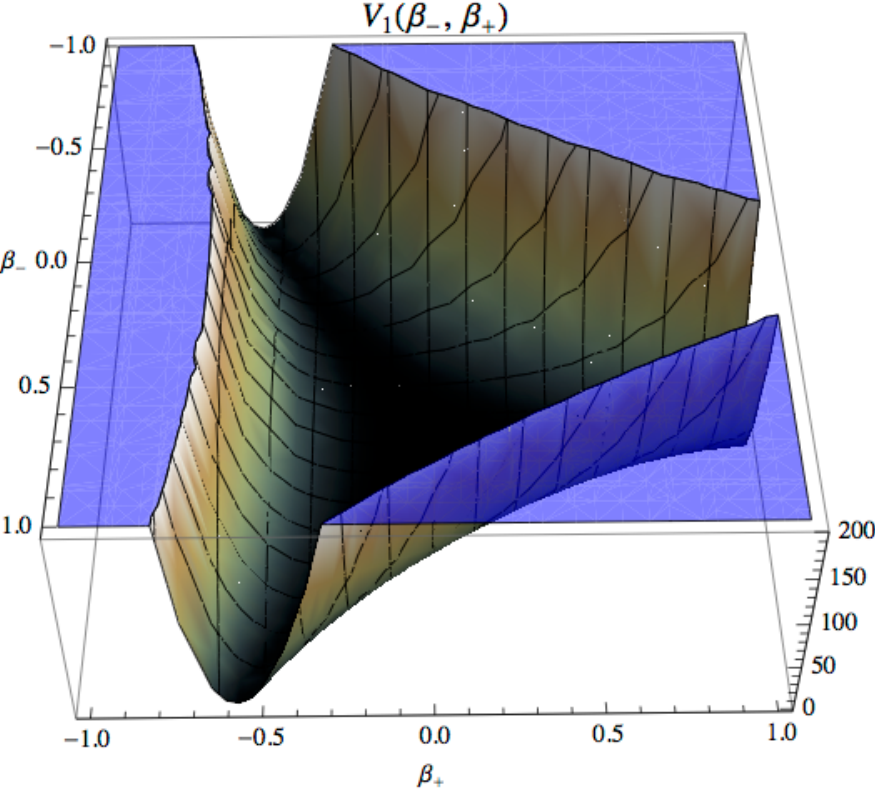}
\caption{The plot of $U_{\mfn} $ for $\mfn=1$ near its minimum. Boundedness from
below,  confining aspects, and $C_{3v}$ symmetry are illustrated.}
\label{Wn}
\end{figure}

The mixmaster universe is a spatially homogenous model with the spatial Killing vector fields satisfying $su(2)$-algebra. It assumes the following metric:
\begin{equation}\label{eq}
\ud s^2= -(24)^2N^2\ud t^2+q^{\frac{4}{3}}\left(
e^{2\beta_++2\sqrt{3}\beta_-}{\omega^1}\otimes{\omega^1}+e^{2\beta_+-2\sqrt{3}\beta_-}{\omega^2}\otimes{\omega^2}+e^{2\beta_+}{\omega^3}\otimes{\omega^3}\right)\, ,
\end{equation}
where $q$, $\beta_+$ and $\beta_-$ are the configuration variables, the $\omega^i$'s are right-invariant dual vectors satisfying
\begin{equation}\label{cartan}
\ud\omega^i=\frac{\mfn}{2}\,\epsilon_{ijk}\,\omega^j\wedge\omega^k,~~\mfn>0\, ,
\end{equation}
and $N$ is the lapse function rescaled in order to simplify coefficients in the subsequent formulas. The fiducial three-volume reads $\mathcal{V}_0=\int\omega^1\wedge\omega^2\wedge\omega^3=\frac{16\pi^2}{\mfn^3}$. We follow the conventions of our previous papers \cite{Bergeron2015short,Bergeron2015long}.  The fiducial volume $\mathcal{V}_0$ of the spatial leaf is chosen to be $\mathcal{V}_0=1$.  Furthermore we are working in a system of units where $16 \pi G c^{-4} = 1=c$. The Hamiltonian reads \cite{Bergeron2015long}:
\begin{equation}
\mathsf{H}= N\left(\frac{9}{4} \, p^2+36\mfn^2q^{2/3} -
\frac{ p_+^2+p_-^2}{q^2}-12 q^{2/3} U_{\mfn}(\beta) \right) \,.
\end{equation}
The potential $U_{\mfn}(\beta)$ involved in the Hamiltonian ${\sf H}$ reads:
\begin{equation}
U_{\mfn}(\beta) = \mfn^2 \frac{e^{4\beta_+}}{3} \left(\left[2\cosh(2\sqrt{3}\beta_-)-e^{- 6\beta_+}
\right]^2-4\right) +  \mfn^2 \,,
\end{equation}
where $\mfn=\sqrt[3]{16\pi^2}$.
This potential possesses the ${\sf C}_{3v}$ symmetry
and  is asymptotically {\it confining} except for the following three
directions (shown in  Fig.\;\ref{Wn}), in which $U_{\mfn} \to 0$:
\begin{equation}\nonumber
(i) ~\beta_-=0, ~\beta_+ \to +\infty,~ (ii)
~\beta_+=- \frac{\beta_-}{\sqrt{3}}, ~\beta_- \to +\infty,~~ (iii)
~\beta_+=  \frac{\beta_-}{\sqrt{3}}, ~\beta_- \to -\infty
\end{equation}
$U_{\mfn}(\beta)$ possesses an absolute minimum for $\beta_\pm=0$,
and near this minimum we have
\begin{equation}
\label{eq:approxupot} U_{\mfn}(\beta) = 8 \mfn^2 (\beta_+^2+\beta_-^2) +
o(\beta_\pm^2) \,.
\end{equation}

\subsection{Quantum Hamiltonian}
\label{Hb9}

The quantization procedure developed in our previous papers \cite{Bergeron2015short, Bergeron2015long}, leads to the following quantum Hamiltonian $ \hat{\mathsf{H}}$
\begin{equation}
\label{quham}
 \hat{\mathsf{H}}=\frac{9}{4} \left(
 \hat{p}^2 + \frac{\hbar^2 \frak{K}_1}{\hat{q}^2} \right) + 36\mfn^2\frak{K}_3 \hat{q}^{2/3} -
 \hat{\mathsf{H}}^{(\mathrm{int})}(\hat{q})\,,
\end{equation}
\begin{equation}
\label{eq:hamilpm}
\hat{\mathsf{H}}^{(\mathrm{int})}(q) :=\frak{K}_2\frac{\hat{p}_+^2+\hat{p}_-^2}{q^2}+
36 \frak{K}_3 q^{2/3}U_{\mfn}(\beta)\,,
\end{equation}
where we made the fundamental split between the isotropic and anisotropic parts of the quantum Hamiltonian. We will refer to anisotropic variables as `internal' degrees of freedom. The $\frak{K}_i$ are purely numerical constants dependent on the choice of the so-called {\it fiducial} vector $\psi$ involved in the quantization procedure. The vector $\psi$ sets a family of coherent states in the Hilbert space of the quantum model. To obtain simplified expressions for the constants, we change our choice of $\psi$ made in our previous paper \cite{Bergeron:2013ika}. We choose another function of rapid decrease on $\mathbb{R}^+$, namely
\begin{equation}
\label{psinu}
\psi_\nu(x) = \left( \frac{\nu}{\pi} \right)^{1/4} \frac{1}{\sqrt{x}} \exp \left[-\frac{\nu}{2} \left(\ln x - \frac{3}{4 \nu} \right)^2  \right] \quad \text{with} \quad \nu >0
\end{equation}
The above function is actually the square root of a Gaussian distribution on the real line with variable $y=\ln x$, centered at $3/4\nu$, and with variance $1/\nu$. With this function we obtain these constants as elementary functions of the free parameter $\nu$
\begin{equation}
\label{Ki}
\frak{K}_1 =  \frac{2 \nu+1}{4}, \quad \frak{K}_2=\exp \left[ \frac{3}{2\nu}\right], \quad \frak{K}_3
=\exp \left[- \frac{1}{18\nu}\right] \,.
\end{equation}
They are also given in Table \ref{ConstK} together with  5 other similar constants whose appearance through various expressions in the article results from our ACS approach.

\begin{table}[t]
  \centering
\begin{tabular}{|c|c|}
    \hline
$\mathfrak{K}_1(\nu)$   & $ \dfrac{2\nu + 1}{4}$   \\
   \hline
  $\mathfrak{K}_2(\nu)$   & $\exp \left[ \dfrac{3}{2\nu}\right]$   \\
   \hline
$\mathfrak{K}_3(\nu)$   & $\exp \left[ -\dfrac{1}{18\nu}\right]$   \\
   \hline
$\mathfrak{K}_4(\nu)$   & $ \left(\nu+ \dfrac{1}{4} \right) \exp \left[ \dfrac{3}{2\nu} \right]$   \\
   \hline
$\mathfrak{K}_5(\nu)$   & $\exp \left[- \dfrac{1}{9 \nu}\right]$   \\
   \hline
$\mathfrak{K}_6(\nu)$   & $\sqrt{2}\,
\exp \left[ \dfrac{1}{\nu}\right]$   \\
   \hline
   $\mathfrak{K}_7(\nu)$   & $\exp \left[ \dfrac{1}{2\nu}\right]$   \\
   \hline
$\mathfrak{K}_8(\nu)$   & $ \exp \left[ \dfrac{3}{2\nu}\right]$   \\
   \hline
\end{tabular}
  \caption{Constants $\mathfrak{K}_i\equiv \mathfrak{K}_i(\nu)$, $i=1,2, \dotsc, 8$ as functions of the free parameter $\nu$ appearing in the fiducial vector \eqref{psinu}}\label{ConstK}
\end{table}

\subsection{Harmonic approximation}
\label{harapp}
In what follows we restrict our considerations to the harmonic approximation of the potential $U_{\mfn}(\beta)$  near its minimum. Therefore the Hamiltonian $\hat{\mathsf{H}}_q$ of Eq. \eqref{eq:hamilpm} reads
\begin{equation}
\label{eq:hqharmonic}
\hat{\mathsf{H}}_q \simeq \frak{K}_2\frac{\hat{p}_+^2+\hat{p}_-^2}{q^2}+
288 \frak{K}_3 \mfn^2 q^{2/3} \left(\hat{\beta}_+^2+ \hat{\beta}_-^2 \right)\,.
\end{equation}

\section{Semiclassical Lagrangian approach}
\label{appsclag}

We recall in this section our procedure detailed in \cite{Bergeron2015long}. It is inspired by Klauder's approach \cite{klauderscm} and is based on a consistent framework allowing to approximate the quantum Hamiltonian and its associated dynamics (in the constraint surface) by making use of the semiclassical Lagrangian approach, which is made possible with the use of our ACS formalism.

\subsection{General setting}

The quantum Hamiltonian \eqref{quham} has the general form (up to constant factors)
\begin{equation}
\label{quhamgen}
 \hat{\mathsf{H}}= \hat{p}^2 + V(\hat{q})
 - \hat{\mathsf{H}}^{(\mathrm{int})}( \hat{q}) \,,
\end{equation}
where the $ q$-dependent Hamiltonian $\hat{\mathsf{H}}^{(\mathrm{int})}(q)$ acts on the Hilbert space of states for internal degrees of freedom, which in our case encode the anisotropy and the shear of the three-geometry.

The Schr\"odinger equation, $$i \hbar \frac{\partial }{\partial t}
| \Psi(t) \rg = N\,\hat{\mathsf{H}} | \Psi(t) \rg$$
can be deduced from the Lagrangian:
\begin{equation}
\label{lagrangen} {\sf L} (\Psi, \dot \Psi, N):=\lg
\Psi(t) |\left( i\hbar\frac{\partial}{\partial t}
-N\,\hat{\mathsf{H}}\right) |\Psi(t)\rg \, ,
\end{equation}
via the minimization of the respective action with respect to $|\Psi(t)\rg$. The quantum counterpart  of the classical constraint
$\mathsf{H}=0$ can be obtained as follows
\begin{equation}
\label{vmH0} \dfrac{\partial {\sf L}}{\partial  N} = \lg
\Psi(t) |\hat{p}^2 + V(\hat{q})-
\hat{\mathsf{H}}^{(\mathrm{int})}( \hat{q})|\Psi(t)\rg=0\, .
\end{equation}
The commonly used Dirac's method of imposing constraints,
$\hat{\mathsf{H}}|\Psi(t)\rg = 0$, implies \eqref{vmH0} but the
reciprocal does not hold in general. This means that a state $|\psi(t)\rg$ for which \eqref{vmH0} holds lies not necessarily in the kernel of the operator $\hat{\mathsf{H}}$.

\subsection{Two kinds of Hamiltonians with internal degrees of freedom}
At this stage, we suppose (due to the confining character of the
potential $U_{\mfn}$) that  $\hat{\mathsf{H}}^{(\mathrm{int})} (q)$ exists as a self-adjoint operator and as a function of the c-number $q$ with a purely discrete spectral decomposition:
\begin{equation}
\label{specdecinq} \hat{\mathsf{H}}^{(\mathrm{int})}(q) = \sum_n
E^{(\mathrm{int})}_n(q) \,
|\phi^{(\mathrm{int})}_n\rg\lg\phi^{(\mathrm{int})}_n|\,.
\end{equation}

To present the Klauder semiclassical procedure in the most general situation (not restricted to cosmology), we distinguish two cases:
\begin{enumerate}
  \item[(I)]   $\phi^{(\mathrm{int})}_n$ is {\it independent} on the external variable $q$, which allows a complete separation of variables, and leads to the original Born-Oppenheimer \cite{born51,sutcliffe12} approach;
  \item[(II)]   $\phi^{(\mathrm{int})}_n$ is {\it dependent} on the external variable $q$.
\end{enumerate}

Different semiclassical approximations are possible leading in particular to the so-called vibronic framework.\footnote{We first present the case (I), which is simple,
and later introduce the more complicated case (II), being applied
to the mixmaster universe.}.

\subsection{Two kinds of semiclassical Lagrangian approximations}
\label{semiclasslagrange}
\subsubsection*{(I) $\phi^{(\mathrm{int})}_n$ independent of $q$}

In this case, a family of exact solutions to the
Schr\"odinger equation $$i\hbar\frac{\partial}{\partial t}
|\Psi(t)\rg = N\,\hat{\mathsf{H}} |\Psi(t)\rg $$ can be introduced in
the form of the tensor product
\begin{equation}
\label{sepvartens} |\Psi(t)\rg = |\phi(t)\rg
\otimes|\phi^{(\mathrm{int})}_n\rg\,,
\end{equation}
where $|\phi(t)\rg$ is a solution to the reduced Schr\"odinger equation
\begin{equation}
\label{tdepredSE} i\hbar\frac{\partial}{\partial t} |\phi(t)\rg
=N\,\hat{\mathsf{H}}^{\mathrm{red}}_n |\phi(t)\rg:=N\left(\hat p^2 + V(\hat{q}) -
E^{(\mathrm{int})}_n( \hat{q}) \right)  |\phi(t)\rg\,
\end{equation}
where $E^{(\mathrm{int})}_n$ is the $n$-th eigenvalue of
$\hat{\mathsf{H}}^{(\mathrm{int})}$ of Eq.\;\eqref{specdecinq}. The $|\Psi(t)\rg $ of Eq.\;\eqref{sepvartens} is a Born-Oppenheimer-type solution.

The equation \eqref{tdepredSE} may be derived from a variational principle applied to the action based on the reduced quantum Lagrangian
\begin{equation}
\label{lagrangian(i)} {\sf L}^{\mathrm{red}}_n (\phi, \dot \phi,N):=\lg \phi(t) | \left(i\hbar\frac{\partial}{\partial
t}
 - N\,\hat{\mathsf{H}}^{\mathrm{red}}_n\right) |\phi(t)\rg \, .
\end{equation}
where the lower index `$n$' indicates the eigenstate of $\hat{\mathsf{H}}^{(\mathrm{int})}$ fixed in the solution.

Following Klauder \cite{klauderscm}, we assume that  $|\phi(t)\rg$ is in fact a (rescaled) affine coherent state $| q(t), p(t) \rg_\lambda $ (see \cite{Bergeron2015long} for more details). The index $\lambda$ in $| q(t), p(t) \rg_\lambda $ used to recall the rescaled character of coherent states will be dropped in the remainder for simplicity.\footnote{We also assume
that the fiducial vector $\psi$ has been chosen in order to obtain the canonical rule $[A_q,A_p]=i \hbar I$ (see\cite{Bergeron2015long}).} Therefore we replace $| \phi(t) \rg$ in ${\sf L}^{\mathrm{red}}_n$ of Eq.\;\eqref{lagrangian(i)} by
\begin{equation}
|  \phi(t) \rg = | q(t), p(t) \rg\,,
\end{equation}
where $q(t)$ and $p(t)$ are some time-dependent functions. Then the Lagrangians  \eqref{lagrangen} and \eqref{lagrangian(i)} turn to assume the semiclassical form
\begin{align}
\label{lagrangian(i)sc} \nonumber {\sf L}^{\mathrm{semi}}_n (q,\dot q,
p, \dot p, N) &= \lg  q(t),p(t) |\left(
i\hbar\frac{\partial}{\partial t} -
N\,\hat{\mathsf{H}}^{\mathrm{red}}_n
\right) |  q(t),p(t)\rg \\ &= - q\dot p -
N\,\lg  q(t),p(t)| \hat{\mathsf{H}}^{\mathrm{red}}_n
|  q(t),p(t)\rg \\
&= - \frac{d}{dt}(qp) + \dot{q} p -
N\,\lg  q(t),p(t)| \hat{\mathsf{H}}^{\mathrm{red}}_n
|  q(t),p(t)\rg\, .
\end{align}
The appearance of the first term $- q\dot p $ in the r.h.s.
of this equation results from the derivative  of coherent states
with respect to parameters $q$ and $p$. Thus, the semiclassical expression for the Hamiltonian is the lower
symbol
\begin{equation}\label{sce}
\mathsf{H}^{\mathrm{semi}}_n (q,p) := \lg  q,p|
\hat{\mathsf{H}}^{\mathrm{red}}_n |  q,p\rg\, .
\end{equation}
It is defined for the `frozen' $n$-th quantum eigenstate of the
internal Hamiltonian, $\hat{\mathsf{H}}^{(\mathrm{int})}(q)$.

Making use of the reduced Hamiltonian \eqref{sce} the equations of motion together with the constraint equation read:
\begin{align}
\label{eqmotred}
  \dot q   & = N\,\frac{\partial}{\partial p}
  \mathsf{H}^{\mathrm{semi}}_n (q,p),\\
   \dot p &= - N\,\frac{\partial}{\partial q}
   \mathsf{H}^{\mathrm{semi}}_n (q,p), \\
  0  &= \mathsf{H}^{\mathrm{semi}}_n (q,p) \,.
\end{align}
These equations will allow us to set up the Friedmann-like equations with quantum corrections for $q$ and $p$.

\subsubsection*{(II)  $\phi^{(\mathrm{int})}_n$ dependent of $q$}

Let us examine the general case in which the eigenstates $|\phi^{(\mathrm{int})}_n\rg$'s depend on $q$. We start again from the spectral decomposition of the internal Hamiltonian
\begin{equation}
\label{specdecinqq} \hat{\mathsf{H}}^{(\mathrm{int})}(q) = \sum_n
E^{(\mathrm{int})}_n(q) \,
|\phi^{(\mathrm{int})}_n(q)\rg\lg\phi^{(\mathrm{int})}_n(q)|\,,
\end{equation}
and we pick some $q$-independent orthonormal basis
$|\tilde{\phi}^{\,(\mathrm{int})}_n\rg$ of the internal Hilbert space
$\mathcal{H}^{(\mathrm{int})}$. This change of basis is associated
with the introduction of the $q$-dependent unitary operator
\begin{equation}
\label{Unchbas} \widetilde{U}(q):= \sum_n |\phi^{(\mathrm{int})}_n(q)\rg\lg
\tilde{\phi}^{\, (\mathrm{int})}_n|\,,
\end{equation}
which allows to deal with the unitary transform of the internal Hamiltonian
\eqref{specdecinq}:
\begin{equation}
\label{specdecinen} \widetilde{\mathsf{H}}^{(\mathrm{int})}(q) =
\widetilde{U}^\dag(q)\hat{\mathsf{H}}^{(\mathrm{int})}(q) \widetilde{U}(q) = \sum_n
E^{(\mathrm{int})}_n(q) \, |\tilde{\phi}^{\,( \mathrm{int})}_n\rg\lg
\tilde{\phi}^{\,(\mathrm{int})}_n|\,.
\end{equation}
The quantum Hamiltonian \eqref{quhamgen} has now the general form
\begin{equation}
\label{quhamgentild}
 \hat{\mathsf{H}}=\hat{p}^2 + V(\hat{q}) - U(\hat q)\widetilde{\mathsf{H}}^{(\mathrm{int})}(\hat q)U^{\dag}
 (\hat q)\,.
\end{equation}
The difference between the Hamiltonians of cases (I) and (II) is
the presence of the unitary operator $U(\hat{q})$ in (II) that
introduces a quantum correlation (entanglement) between the internal degrees of freedom and the external one. As a consequence, any solution $|\Psi(t)\rg$ of the Schr\"odinger equation $i\hbar\frac{\partial}{\partial t} |\Psi(t)\rg = N\,\hat{\mathsf{H}}|\Psi(t)\rg $ cannot be factorized as a tensor product like $|\phi(t)\rg \otimes|\phi^{(\mathrm{int})}(t)\rg$, contrarily to
the case (I). In our case we wish to follow Klauder's approach to
build some semiclassical Lagrangian analoguous to
Eq.\;\eqref{lagrangian(i)}. We use the previous case (I) as a
reference pattern to build approximate possible forms of
$|\Psi(t)\rg$.\\

Let us introduce the $q$-dependent operator
$\hat{\mathrm{A}}(q)$ acting on the internal Hilbert space
$\mathcal{H}^{(\mathrm{int})}$
\begin{equation}
\label{gaugefield} \hat{\mathrm{A}}(q) = i \hbar \frac{\ud \widetilde{U}}{\ud
q}(q) \widetilde{U}^\dag (q) \,.
\end{equation}
As a matter of fact, $\hat{\mathrm{A}}(q)$ is self-adjoint. We define the unitary transform of the total Hamiltonian $\hat{\mathsf{H}}$ of Eq.\;\eqref{quhamgentild}
\begin{equation}\label{UredefHam}
\tilde{\mathsf{H}}= \widetilde{U}^{\dag}(\hat q)\hat{\mathsf{H}}\widetilde{U}(\hat q)
\end{equation}
and we find
\begin{equation}
\label{quhamgentransform} \tilde{\mathsf{H}} = (\hat{p} - \hat{\mathrm{A}}(\hat{q}))^2 +
V(\hat{q}) - \widetilde{\mathsf{H}}^{(\mathrm{int})}(\hat
q)\,.
\end{equation}
Since the eigenvectors $|\tilde{\phi}^{\,( \mathrm{int})}_n\rg$ of $\widetilde{\mathsf{H}}^{(\mathrm{int})}$ do not depend on $q$, the Hamiltonian $\tilde{\mathsf{H}}$ resembles the case (I) but with a supplementary potential $\hat{\mathrm{A}}$, which couples internal and external degrees of freedom. How to proceed in this case is discussed in the next section.

We note that the Hamiltonian $\tilde{\mathsf{H}}$ of Eq.\;\eqref{quhamgentransform} is the most general form of the Hamiltonian of a particle moving on a half-line and minimally coupled to an external field. It is also the most general form, which admits the ``shadow Galilean invariance''  \cite{JMLL}. In other words, at any fixed time a state of this interacting particle is indistinguishable from a state of a free particle and obeys the Galilean addition rule of velocities.

\section{Discussion of the framework}
\label{discframe}
We take into account the fact that the Hamiltonian of Eq.\;\eqref{quhamgentild}, belonging to the case (II), can be unitarily transformed into the Hamiltonian of Eq.\;\eqref{quhamgentransform}, which resembles the case (I). We build on the analysis of the previous case (I) and list various possible ansatzes of $| \Psi(t) \rg$:

\begin{enumerate}
  \item[(a)] In the first approach we keep the tensor product expression
  of (I), but inserting the $q$-dependence of eigenstates. This corresponds
to a Born-Oppenheimer-like approximation:
  \begin{equation}
\label{firstapp} |\Psi(t)\rg \approx |  q(t),p(t)\rg\otimes
|\phi^{(\mathrm{int})}_n(q(t))\rg \,.
\end{equation}
  \item[(b)] The second strategy consists in introducing some (minimal)
  entanglement between $q$ and internal degrees of freedom.
  This corresponds to a Born-Huang-like approximation:
   \begin{equation}
\label{secapp} |\Psi(t)\rg \approx \widetilde{U}(\hat q)
\left(|  q(t),p(t)\rg\otimes | \tilde{\phi}^{\,(\mathrm{int})}_n\rg\right)\,.
\end{equation}
  \item[(c)] In the third method one keeps the tensor product approximation,
  but employs a general time-dependent state for the internal degrees of
  freedom:
   \begin{equation}
\label{thirdapp} |\Psi(t)\rg \approx |  q(t),p(t)\rg\otimes
|\phi^{(\mathrm{int})}(t)\rg \,.
\end{equation}
  \item[(d)] The fourth strategy is the most  general one.  It  consists
  in merging (b) and
  (c):
   \begin{equation}
\label{fourthapp} |\Psi(t)\rg \approx \widetilde{U}(\hat q) \left(
| q(t),p(t)\rg\otimes  |\tilde{\phi}^{\, (\mathrm{int})}(t)\rg \right) \,.
\end{equation}
\end{enumerate}

Building the semiclassical Lagrangian in agreement with the
procedure defined in (I), we can distinguish two categories in the
approximations listed above.
\begin{itemize}
   \item[(1)] (a) and (b) are completely manageable on the
    semiclassical level: they involve $q$ and $p$ as dynamical
    variables, while the anisotropy degrees of freedom are `frozen' in
    some eigenstate; (a) and (b) correspond typically to the adiabatic
    approximations.
\item[(2)] (c) and (d) are more elaborate: they
mix a semiclassical dynamics for $(q,p)$ and a quantum dynamics
for the anisotropy degrees of freedom. This corresponds to
the vibronic approximations, well-known in molecular physics and quantum chemistry \cite{yakorny12}. In our case this means that different quantum eigenstates of the
anisotropy degrees of freedom are involved in the dynamics: during
the evolution excitations and decays are possible, with an
exchange of energy with  the `classical degree of freedom'
$(q,p)$.
\end{itemize}

The Bianchi-IX Hamiltonian belongs to the general case (II). In our previous papers \cite{Bergeron2015short,Bergeron2015long} we restrict ourselves to the analysis of the simplest cases (a) and (b). Presently we investigate the most involved case (d) corresponding to the vibronic-like approximation. We assume the harmonic approximation to the anisotropy potential $U_{\mfn}$ in order to obtain analytical expressions of the dynamical equations.

\section{Dynamical equations}
\label{dyneq}
\subsection{Semiclassical Lagrangian}
\label{sclag}
In the case of the harmonic approximation of $U_{\mfn}$, the internal Hamiltonian $\hat{\mathsf{H}}^{(\mathrm{int})}(q)$ of our cosmological model reads
\begin{equation}
\label{eq:hqharmonic1}
\hat{\mathsf{H}}^{(\mathrm{int})}(q) :=\frak{K}_2\frac{\hat{p}_+^2+\hat{p}_-^2}{q^2}+
288 \frak{K}_3 \mfn^2 q^{2/3} \left(\hat{\beta}_+^2+ \hat{\beta}_-^2 \right)\,.
\end{equation}
We define the operator $\widetilde{U}(q)$ as the unitary dilations for the anisotropic variables:
\begin{equation}
\widetilde{U}(q) = e^{\frac{2 i}{3} (\ln q) \, \hat{\mathrm{D}}}
\,,
\end{equation}
with
\begin{equation}
\hat{\mathrm{D}} = \hat{\mathrm{D}}_+ + \hat{\mathrm{D}}_- \, ,
\quad \hat{\mathrm{D}}_\pm = \frac{1}{2 \hbar}(\hat{p}_\pm
\hat{\beta}_\pm + \hat{\beta}_\pm \hat{p}_\pm) \,.
\end{equation}
In agreement with the notation in \eqref{specdecinen} we have
\begin{equation}
\hat{\mathsf{H}}^{(\mathrm{int})}(q) = \widetilde{U}(q) \widetilde{\mathsf{H}}^{(\mathrm{int})}(q) \widetilde{U}^\dag(q)\,,
\end{equation}
where
\begin{equation}
\widetilde{\mathsf{H}}^{(\mathrm{int})}(q) =\frac{1}{q^{2/3}} \left( \frak{K}_2 \left(\hat{p}_+^2+\hat{p}_-^2 \right)+
288 \frak{K}_3 \mfn^2 \left(\hat{\beta}_+^2+ \hat{\beta}_-^2 \right) \right)\,.
\end{equation}
We notice that $\widetilde{\mathsf{H}}^{(\mathrm{int})}(q)$ is the usual harmonic Hamiltonian multiplied by a $q$-dependent factor and thus, it is diagonal in a fixed, $q$-independent basis as needed. Moreover, it possesses the normal form
\begin{equation}
\label{hintnormal}
\widetilde{\mathsf{H}}^{(\mathrm{int})}(q) = \frac{24 \hbar \sqrt{2 \frak{K}_2 \frak{K}_3} \mfn}{q^{2/3}} \left( \hat{{\sf N}}_+ + \hat{{\sf N}}_-+1 \right)\,,
\end{equation}
where $\hat{{\sf N}}_\pm = \hat{a}^\dag_\pm \, \hat{a}_\pm$ are the number operators and $\hat{a}_\pm$ and $\hat{a}^\dag_\pm$ are the lowering and raising operators. We introduce the harmonic basis $|\tilde{\phi}^{\,(\mathrm{int})}_{n_+,n_-}\rg$ such that
\begin{align}
\hat{{\sf N}}_\pm \, |\tilde{\phi}^{\,(\mathrm{int})}_{n_+,n_-}\rg &= n_\pm \, |\tilde{\phi}^{\,(\mathrm{int})}_{n_+,n_-}\rg\\
\hat{a}_+ \, |\tilde{\phi}^{\,(\mathrm{int})}_{n_+, n_-}\rg &= \sqrt{n_+} \, |\tilde{\phi}^{\,(\mathrm{int})}_{n_+-1,\, n_-}\rg\\
\hat{a}_- \, |\tilde{\phi}^{\,(\mathrm{int})}_{n_+, n_-}\rg &= \sqrt{n_-} \, |\tilde{\phi}^{\,(\mathrm{int})}_{n_+,\, n_--1}\rg \,.
\end{align}
The operators $\hat{p}_\pm$ and $\hat{\beta}_\pm$ read
\begin{align}
\hat{\beta}_\pm &= \frac{1}{4} \sqrt{\frac{2\hbar}{3 \mfn}} \left( \frac{\frak{K}_2}{2 \frak{K}_3} \right)^{1/4} (\hat{a}_\pm + \hat{a}^\dag_\pm)\\
\hat{p}_\pm &= i \sqrt{6 \hbar \mfn} \left( \frac{2\frak{K}_3}{ \frak{K}_2} \right)^{1/4} (\hat{a}^\dag_\pm - \hat{a}_\pm)\,,
\end{align}
while
\begin{equation}
\hat{\mathrm{D}}_\pm = \frac{i}{2} \left((\hat{a}^\dag_\pm)^2 - (\hat{a}_\pm)^2 \right)\,.
\end{equation}
The `gauge field' $\hat{\mathrm{A}}(q)$ defined
in Eq.\;\eqref{gaugefield}, is specified as
\begin{equation}
\hat{\mathrm{A}}(q) = - \frac{2 \hbar}{3 q} \hat{\mathrm{D}} \,,
\end{equation}
and the Hamiltonian $\hat{\mathsf{H}}$ of Eq.\;\eqref{quham}, transformed accordingly to Eq.\;\eqref{UredefHam}, reads
\begin{equation}
\label{vibrohamil}
\tilde{\mathsf{H}} = \frac{9}{4} (\hat{p} - \hat{\mathrm{A}}(\hat{q}))^2 +
\frac{\hbar^2 \frak{K}_1}{\hat{q}^2} + 36\mfn^2\frak{K}_3 \hat{q}^{\frac{2}{3}}- \widetilde{\mathsf{H}}^{(\mathrm{int})}(\hat
q) \,.
\end{equation}
where $\widetilde{\mathsf{H}}^{(\mathrm{int})}(q)$ is given in Eq.\;\eqref{eq:hqharmonic1}. Now assuming that the quantum states $| \Psi(t) \rg$ correspond to the case (d) presented above  (i.e. correspond to Eq.\;\eqref{fourthapp}), we have
   \begin{equation}
   \label{psivibronic}
 |\Psi(t)\rg = \widetilde{U}(\hat q) \left(
| q(t),p(t)\rg\otimes  |\tilde{\phi}^{\,(\mathrm{int})}(t)\rg \right) \,,
\end{equation}
where the  $|\tilde{\phi}^{\,(\mathrm{int})}(t)\rg$ are assumed to be normalized and
\begin{equation}
|\tilde{\phi}^{\,(\mathrm{int})}(t)\rg = \sum_{n_+,n_-} c_{n+,n_-}(t) |\tilde{\phi}^{\,(\mathrm{int})}_{n_+,n_-}\rg\,.
\end{equation}
The expectation value
$$\lg \Psi(t) \,| \hat{\mathsf{H}}  | \Psi(t) \rg = \left(\lg \tilde{\phi}^{\,(\mathrm{int})}(t) \,|\otimes\lg q(t),p(t) \,|\right) \tilde{\mathsf{H}} \left( | q(t),p(t) \rg\otimes  |\tilde{\phi}^{\,(\mathrm{int})}(t)\rg\right)$$
can be expanded as
\begin{equation}
\lg \Psi(t) \,| \hat{\mathsf{H}}  | \Psi(t) \rg ={\sf H}^{\mathrm{semi}}_{E}(q(t),p(t))- \lg \tilde{\phi}^{\,(\mathrm{int})}(t)| \, \widetilde{\mathsf{H}}_{I}^{\mathrm{semi}}(q(t),p(t)) \, | \tilde{\phi}^{\,(\mathrm{int})}(t) \rg\,,
\end{equation}
where ${\sf H}^{\mathrm{semi}}_E(q,p)$ is a pure $c$-number (real function of $(q,p)$) and the new quantum Hamiltonian $\widetilde{\mathsf{H}}_I^{\mathrm{semi}}(q,p)$ is an operator that depends parametrically on $(q,p)$ and acts on the Hilbert space of quantized anisotropy degrees of freedom. Note that the upper index `$\mathrm{semi}$' refers only to the isotropic variables. It follows that
\begin{align}
{\sf H}^{\mathrm{semi}}_E(q,p) & =\frac{9}{4} \left[ p^2+ \frac{\hbar^2 \frak{K}_4}{q^2} \right] + 36 \mfn^2 \frak{K}_5 \, q^{2/3}- \frac{24 \hbar}{q^{2/3}} \frak{K}_6 \, \mfn \\
\widetilde{\mathsf{H}}_I^{\mathrm{semi}}(q,p) &= \frac{24 \hbar}{q^{2/3}} \frak{K}_6 \, \mfn \, \left(\hat{{\sf N}}_++\hat{{\sf N}}_- \right) -\frac{\hbar^2 \frak{K}_8
}{q^2} \,  \hat{{\sf D}}^2 - \frac{3\hbar}{q} \frak{K}_7 \, p \, \hat{{\sf D}} \,.
\end{align}
The positive numerical constants $\frak{K}_4$,  $\frak{K}_5$, $\frak{K}_6$, $\frak{K}_7$, $\frak{K}_8
$, are given as functions of the free fiducial parameter $\nu$ in Table \ref{ConstK}.

Using the ansatz for $| \Psi(t) \rg$ specified in Eq.\;\eqref{psivibronic}, the first part of the semiclassical Lagrangian corresponding to Eq. \eqref{lagrangen} is expanded as
\begin{equation}
i \hbar \, \lg \Psi(t) | \frac{\partial}{\partial t} | \psi(t) \rg = -q\, \dot{p} + i \hbar \, \lg \tilde{\phi}^{\,(\mathrm{int})}(t) | \frac{\partial}{\partial t} | \tilde{\phi}^{\,(\mathrm{int})}(t) \rg \,.
\end{equation}
Therefore the semiclassical Lagrangian ${\sf L}^{\mathrm{semi}}$ involving both the classical variables $(q,p)$ and the quantum ones $|\tilde{\phi}^{\,(\mathrm{int})}(t) \rg$ finally reads
\begin{align}\label{finallagrangian}
{\sf L}^{\mathrm{semi}}(q,\dot{q},p,\dot{p},\tilde{\phi}^{\,(\mathrm{int})},\dot{\tilde{\phi}}^{\,(\mathrm{int})},N) &=-q\, \dot{p} + i \hbar \, \lg \tilde{\phi}^{\,(\mathrm{int})}(t) | \frac{\partial}{\partial t} | \tilde{\phi}^{\,(\mathrm{int})}(t) \rg  - N\, {\sf H}^{\mathrm{semi}}_E(q,p)\\
\notag &+ N\, \lg \tilde{\phi}^{\,(\mathrm{int})}(t) | \, \widetilde{\mathsf{H}}_I^{\mathrm{semi}}(q,p) \, |\tilde{\phi}^{\,(\mathrm{int})}(t) \rg \,.
\end{align}

\subsection{Complete set of dynamical equations}
\label{dyneqlag}

\subsubsection{Reparametrization-invariant evolution}
By the minimization of the action of Eq.\;\eqref{finallagrangian}, we obtain the complete set of dynamical equations of our system, together with the semiclassical constraint:
\begin{align}
\label{dyn1} \dot{q}&=  N\, \partial_p {\sf H}^{\mathrm{semi}}_E(q,p) -N\,  \lg \tilde{\phi}^{\,(\mathrm{int})} | \, \partial_p \,\widetilde{\mathsf{H}}_I^{\mathrm{semi}}(q,p) \, | \tilde{\phi}^{\,(\mathrm{int})} \rg\\
\label{dyn2} \dot{p}&= - N\, \partial_q {\sf H}^{\mathrm{semi}}_E(q,p) +N\, \lg \tilde{\phi}^{\,(\mathrm{int})} | \, \partial_q \,\widetilde{\mathsf{H}}_I^{\mathrm{semi}}(q,p) \, | \tilde{\phi}^{\,(\mathrm{int})} \rg\\
\label{dyn3} - i \hbar \,\partial_t  &\, | \phi^{(\mathrm{int})}(t) \rg = N \, \hat{\mathsf{H}}_I^{\mathrm{semi}}(q,p) \, | \phi^{(\mathrm{int})}(t) \rg\\
\label{dyn4} \partial_{N} {\sf L}^{\mathrm{semi}} &= {\sf H}^{\mathrm{semi}}_E(q,p) -\lg \tilde{\phi}^{\,(\mathrm{int})} | \, \widetilde{\mathsf{H}}_I^{\mathrm{semi}}(q,p) \, | \tilde{\phi}^{\,(\mathrm{int})} \rg = 0
\end{align}
We notice in Eq. \eqref{dyn3} the minus sign in ``$-i \hbar \partial_t \dots$" due to the choice of sign in the definition of $\widetilde{{\sf H}}_I^{\mathrm{semi}}$.

Eqs. \eqref{dyn1}, \eqref{dyn2}, \eqref{dyn3} are found to read explicitly
\begin{align}
\label{dynp1} N^{-1}\,  \dot{q} =& \frac{9}{2} p+\frac{3 \hbar}{q} \frak{K}_7 \, \lg \hat{{\sf D}} \rg \\
\label{dynp2} N^{-1}\,  \dot{p} =& \frac{3 \hbar}{q^2} \frak{K}_7 \lg \hat{{\sf D}} \rg \, p + \frac{9 \hbar^2}{2 q^3} \left[ \frak{K}_4 + \frac{4}{9} \frak{K}_8
 \lg \hat{{\sf D}}^2 \rg \right] - 24\mfn^2 \frac{\frak{K}_5}{q^{1/3}} \\
\notag &-\frac{48 \hbar}{3 q^{5/3}} \frak{K}_6 \mfn \, \lg \hat{{\sf N}}_+ + \hat{{\sf N}}_- + 1 \rg \\
\label{dynp3} -i \, q^{2/3} \,N^{-1}\,  \partial_t \, | \tilde{\phi}^{\,(\mathrm{int})} \rg =& \left( 24 \frak{K}_6 \mfn \left( \hat{{\sf N}}_+ + \hat{{\sf N}}_- \right) -\frac{3 \frak{K}_7}{q^{1/3}} \, p \, \hat{{\sf D}} - \frac{\hbar \frak{K}_8
}{q^{4/3}} \, \hat{{\sf D}}^2 \right) \, |\tilde{ \phi}^{\,(\mathrm{int})} \rg
\end{align}
where $\lg \, . \, \rg \equiv \lg \tilde{\phi}^{\,(\mathrm{int})} | \, . \, | \tilde{\phi}^{\,(\mathrm{int})} \rg$.

The dynamical equations \eqref{dynp1}, \eqref{dynp2}, \eqref{dynp3} complemented with the semiclassical constraint \eqref{dyn4} constitute the vibronic-like framework for our system.
The choice of a finite linear combination of eigenstates for $| \tilde{\phi}^{\,(\mathrm{int})}(t) \rg$ like
\begin{equation}
| \tilde{\phi}^{\,(\mathrm{int})}(t) \rg = \sum_{n_+=0,n_-=0}^N c_{n+,n_-}(t) |\tilde{\phi}^{\,(\mathrm{int})}_{n_+,n_-}\rg\,
\end{equation}
allows a numerical integration of the dynamical equations to obtain the time behavior of $q(t)$, $p(t)$ and $c_i(t)$.

\subsubsection{Conformal time evolution}
We parametrize the trajectory with the rescaled conformal time, which is denoted by `$\eta$' and defined as $24 a\ud \eta= \ud t$, where `$t$' is the cosmological time\footnote{The coefficient $24$ appears to due to our definition of the rescaled lapse function in Eq. (\ref{eq}).}. Therefore we set the lapse function as
\begin{equation}
 N=q^{2/3} \,.
\end{equation}
We choose as new conjugate variables $(a,p_a)$, the scale factor and its associated momentum,
\begin{equation}
\label{apa}
a=q^{2/3}\, , \qquad p_a = (3/2) q^{1/3} p\, .
\end{equation}
Taking into account the constraint \eqref{dyn4}, the dynamical equations \eqref{dynp1}, \eqref{dynp2}, \eqref{dynp3} become
\begin{align}
\label{dyns1}  \frac{\ud}{\ud \eta}  \, a = & \, 2 \left( p_a + \frac{\hbar \frak{K}_7}{a}  \lg \hat{{\sf D}} \rg  \right)\\
\label{dyns2}  \frac{\ud}{\ud \eta}  \, p_a =& \, \frac{2 \hbar}{a^2} \frak{K}_7  \lg \hat{{\sf D}} \rg \, p_a - 60 \mfn^2 \frak{K}_5 \, a + \frac{9 \hbar^2}{2 a^3} \left( \frak{K}_4 + \frac{4}{9} \frak{K}_8
 \lg  \hat{{\sf D}}^2 \rg\right) \\
\label{dyns3} - i \,  \frac{\ud}{\ud \eta}  | \tilde{\phi}^{(\mathrm{int})} \rg = & \left[ 24 \frak{K}_6 \mfn \left(\hat{{\sf N}}_+ + \hat{{\sf N}}_-\right)-\frac{2 \frak{K}_7}{a} p_a \hat{{\sf D}} -\frac{\hbar \frak{K}_8
}{a^2} \hat{{\sf D}}^2 \right] \,  |\tilde{\phi}^{(\mathrm{int})} \rg
\end{align}
where $\lg \, . \, \rg \equiv \lg \tilde{\phi}^{(\mathrm{int})} | \, . \, | \tilde{\phi}^{(\mathrm{int})} \rg$.
From Eq. \eqref{dyns3} we see  that in the absence of the coupling terms in $\hat{{\sf D}}$ and $\hat{{\sf D}}^2$, the quantum degrees of freedom evolve freely in the conformal time, i.e. with the Hamiltonian $ 24 \frak{K}_6 \mfn \left(\hat{{\sf N}}_+ + \hat{{\sf N}}_-\right)$. This situation corresponds to the adiabatic approximation studied in our previous papers  \cite{Bergeron2015short, Bergeron2015long}.

\subsection{Modified Friedmann equation}
\label{modfried}
The constraint of Eq.\;\eqref{dyn4} reads
\begin{align}
\label{dyns4} \left(p_a+\frac{\hbar \frak{K}_7}{a}  \lg \hat{{\sf D}} \rg \right)^2 + & \frac{\hbar^2}{a^2} \left[ \frac{9 \frak{K}_4}{4} + \frak{K}_8
  \lg \hat{{\sf D}}^2 \rg - \frak{K}_7^2 \lg \hat{{\sf D}} \rg^2 \right] + 36 \mfn^2 \frak{K}_5 a^2\\
\notag & = 24 \hbar \frak{K}_6 \mfn \left( \lg \hat{{\sf N}}_+ + \hat{{\sf N}}_- \rg +1 \right)
\end{align}
Let us notice that
\begin{equation}
\frak{K}_8
  \lg \hat{{\sf D}}^2 \rg - \frak{K}_7^2 \lg \hat{{\sf D}} \rg^2 = \frak{K}_8
 (\Delta \hat{{\sf D}})^2 +(\frak{K}_8
 - \frak{K}_7^2) \lg \hat{{\sf D}} \rg^2.
\end{equation}
Since it can be verified that $\frak{K}_8 \ge \frak{K}_7^2$, we deduce
\begin{equation}
\frak{K}_8
  \lg \hat{{\sf D}}^2 \rg - \frak{K}_7^2 \lg \hat{{\sf D}} \rg^2 \ge 0 \,.
\end{equation}
Furthermore, defining the Hubble rate as $\theta:=N^{-1} \dot{a}/a$, it reads with the conformal time derivative
\begin{equation}
\theta = \frac{1}{a^2}  \frac{\ud a}{\ud \eta} \,.
\end{equation}
Therefore, using Eq. \eqref{dyns1}, we obtain
\begin{equation}
\theta = \frac{2}{a^2} \left( p_a+ \frac{\hbar \frak{K}_7}{a} \lg \hat{{\sf D}} \rg \right) \,.
\end{equation}
Eq.\;\eqref{dyns4} can be rewritten as the modified Friedmann equation
\begin{equation}
\label{friedmod}
\theta^2 + \frac{4 \hbar^2}{a^6} \left[ \frac{9 \frak{K}_4}{4} + \frak{K}_8
  \lg \hat{{\sf D}}^2 \rg - \frak{K}_7^2 \lg \hat{{\sf D}} \rg^2 \right] + \frac{144 \mfn^2 \frak{K}_5}{ a^2}
= \frac{96 \hbar \frak{K}_6 \mfn}{a^4} \left( \lg \hat{{\sf N}}_+ + \hat{{\sf N}}_- \rg +1 \right)
\end{equation}
We recover the modified Friedman equation of our previous paper \cite{Bergeron2015long}. But let us notice that in \cite{Bergeron2015long} the expectation values $\lg \hat{{\sf D}} \rg$, $\lg \hat{{\sf D}}^2 \rg$ and $\lg \hat{{\sf N}}_+ + \hat{{\sf N}}_- \rg$ are all time-independent due to the adiabatic approximation, i.e. due to the stationary behavior of $| \tilde{\phi}^{(\mathrm{int})} (t) \rg$ in the Born-Huang approximation. Now these quantities become time-dependent as their evolution is governed by the Schr\"{o}dinger equation \eqref{dyns3}. Therefore a simple algebraic analysis of Eq. \eqref{friedmod} is not sufficient to conclude about singularity avoidance, contrary to the case considered in \cite{Bergeron2015long}.

\section{Comments on the dynamical equations}
\label{comdyneq}

\subsection{Dynamics}
The equations \eqref{dyns1}, \eqref{dyns2}, \eqref{dyns3} and \eqref{dyns4} constitute a complete self-consistent dynamical framework, in which the isotropic degree of freedom (scale factor) $a$ is semiclassical and the anisotropy degrees of freedom $\beta_\pm$ are quantum. The equations \eqref{dyns1}, \eqref{dyns2}, \eqref{dyns3} include both the dynamical action of $a$ on the quantized $\beta_\pm$, and the backreaction of the quantized $\beta_\pm$ on $a$.

\subsection{Constraint}
In this model the kinematical phase-space $\mathcal{P}$ of the system is a mixture of the classical phase-space, $\R_+ \times \R$, and the quantum one, the internal Hilbert space $\mathcal{H}^{(\mathrm{int})}$, namely $$\mathcal{P} = \R_+ \times \R \times \mathcal{H}^{(\mathrm{int})}\;.$$ The constraint equation \eqref{dyn4} (or, \eqref{dyns4}) neither completely falls  within the usual classical domain (i.e. the hypersurface in $\R^{2n}$ fixed by the classical constraint equation) nor within the quantum domain (i.e. the kernel of a quantum constraint operator, if we adopt the Dirac point of view). The minimization of the action of Eq.\;\eqref{finallagrangian} with respect to the lapse function $N$  gives the most consistent way to address this problem: the physical states form some sub-manifold of $\mathcal{P}$.

\subsection{Time covariance}
As the equations \eqref{dyn1}, \eqref{dyn2}, \eqref{dyn3} and \eqref{dyn4} originate from the time-reparametrization invariant semiclassical action, they naturally exhibit invariance with respect to a change of time parameter $t \mapsto t'(t,q,p)$. This is made apparent via keeping the lapse function $N=N(t,q,p)>0$ undetermined in three of the four equations. Therefore, as within the classical framework, there is no preferred time. Note that the time reparametrization-invariance holds for both the semiclassical and quantum degrees of freedom.

\subsection{Adiabaticity}
The adiabatic situation corresponds to the case when the initial state $| \tilde{\phi}^{(\mathrm{int})}(t=0) \rg$ is chosen as an eigenstate of $\tilde{{\sf H}}^{(\mathrm{int})}$ of Eq. \eqref{hintnormal}, i.e. $| \phi^{(\mathrm{int})}(t=0) \rg = |\phi^{(\mathrm{int})}_{n_+,n_-}\rg$, and is assumed stationary during time evolution. Obviously this situation is only an approximation of the real case, in which the anisotropy quantum state evolves accordingly to Eq.\;\eqref{dyns3}. Therefore, the possible excitations or decays and their backreaction induced by the complete set of dynamical equations \eqref{dyns1}, \eqref{dyns2}, \eqref{dyns3} represent the sought nonadiabatic dynamical effects.

\subsection{Bounce and re-collapse}
For the adiabatic case \cite{Bergeron2015long}, we have proved that the trajectory of the scale factor is dynamically complete and at low volumes exhibits a bounce due to the repulsive potential $ \propto a^{-6}$ present in the modified Friedmann equation \eqref{friedmod}. Moreover, it is closed for $\nu \in [0, \nu_m \simeq7.4]$ and the re-bounce taking place for large values of $a$ is due to the term $\propto\frak{K}_5 a^{-2}$.

In what follows we want  to analyze the nonadiabatic quantum effects near the bounce (if the bounce exists within the general nonadiabatic case studied herein). The re-collapse is not interesting in this framework. Furthermore, the re-collapse in the real context also depends on the matter content of the universe which is not taken into account presently. Therefore in what follows we assume $\frak{K}_5=0$ to remove any re-recollapse. A single bounce joining a contracting phase to an expanding one may exist in this case.

\section{Numerical simulations}
\label{numsim}
\subsection{Simplified framework}
\label{simframe}
Let us introduce a simplified model involving a single anisotropy degree of freedom specified by a unique quantum number operator $\hat{{\sf N}}=\hat{a}^\dagger \, \hat{a}$. We introduce the rules of replacement $\hat{{\sf N}}_++\hat{{\sf N}}_- \mapsto 2 \hat{{\sf N}}$ and $\hat{{\sf D}}_++\hat{{\sf D}}_- \mapsto 2 \hat{{\sf D}}$ where now
\begin{equation}
\hat{\mathrm{D}} = \frac{i}{2} \left((\hat{a}^\dag)^2 - (\hat{a})^2 \right)\,.
\end{equation}
 The dynamical equations \eqref{dyns1}, \eqref{dyns2} and \eqref{dyns3} become
\begin{align}
\label{dynss1}  \frac{\ud}{\ud \eta}  \, a = & \, 2 \left( p_a + \frac{2 \hbar \frak{K}_7}{a}  \lg \hat{{\sf D}} \rg  \right)\\
\label{dynss2}  \frac{\ud}{\ud \eta}  \, p_a =& \, \frac{4 \hbar}{a^2} \frak{K}_7  \lg \hat{{\sf D}} \rg \, p_a - 60 \mfn^2 \frak{K}_5 \, a + \frac{9 \hbar^2}{2 a^3} \left( \frak{K}_4 + \frac{16}{9} \frak{K}_8
 \lg  \hat{{\sf D}}^2 \rg\right) \\
\label{dynss3} - i \,  \frac{\ud}{\ud \eta}  | \tilde{\phi}^{\,(\mathrm{int})} \rg = & \left[ 48 \frak{K}_6 \mfn \, \hat{{\sf N}} -\frac{4 \frak{K}_7}{a} p_a \hat{{\sf D}} -\frac{4 \hbar \frak{K}_8
}{a^2} \hat{{\sf D}}^2 \right] \,  | \tilde{\phi}^{\,(\mathrm{int})} \rg
\end{align}
where $\lg \, . \, \rg \equiv \lg \tilde{\phi}^{\,(\mathrm{int})} | \, . \, | \tilde{\phi}^{\,(\mathrm{int})} \rg$. The constraint of Eq. \eqref{dyns4} reads
\begin{align}
\label{dynss4} \left(p_a+\frac{2\hbar \frak{K}_7}{a}  \lg \hat{{\sf D}} \rg \right)^2 + & \frac{\hbar^2}{a^2} \left[ \frac{9 \frak{K}_4}{4} + 4\frak{K}_8
  \lg \hat{{\sf D}}^2 \rg - 4\frak{K}_7^2 \lg \hat{{\sf D}} \rg^2 \right] + 36 \mfn^2 \frak{K}_5 a^2\\
\notag & = 24 \hbar \frak{K}_6 \mfn \left( \lg 2 \hat{{\sf N}} \rg +1 \right)
\end{align}

We perform numerical simulations by taking the following steps:
\begin{enumerate}
\item We choose $\nu=1.7$ corresponding to the minimal value of the constant $\mathfrak{K}_4(\nu)$, leading to the maximal possible excitations as found in our previous analysis \cite{B9vibronica}. The constant $\mfn$ is fixed as $\mfn=\sqrt[3]{16\pi^2}$. We work in a system of units where $\hbar=1$ and we impose $\frak{K}_5=0$ to avoid the re-collapse.

\item Then we choose some initial condition for the triplet $(a_0,p_{a,0}, |\tilde{\phi}^{\,({\rm int})}_0 \rangle)$ compatible with the constraint of Eq. \eqref{dynss4}.

\item Once $a_0$ and $|\tilde{\phi}^{\,({\rm int})}_0 \rg$ have been  specified, the constraint of Eq. \eqref{dynss4} allows to determine $p_{a,0}$ up to a sign. We impose $p_{a,0}<0$ to start with a contracting phase and to study the effects of bounce. We choose $a_0=5$ to start the integration far from the singularity.

\item We select various initial states for the anisotropy degree of freedom $|\tilde{\phi}^{\,({\rm int})}_0 \rangle=|n_0 \rangle$ with $n_0=0,1,2,\dots$ and we perform a numerical integration of Eqs \eqref{dynss1}, \eqref{dynss2} and \eqref{dynss3}. The states $|\tilde{\phi}^{\,({\rm int})} \rangle (\eta)$ are approximated by a finite combination $|\tilde{\phi}^{\,({\rm int})} \rangle(\eta) =\sum_n c_n(\eta) \, |n \rg$, which consists of a sufficiently large number of eigenstates $|n \rg$.

\item We determine the evolution in the conformal time of the scale factor $a(\eta)$, of the populations $|c_n(\eta)|^2=| \lg n | \tilde{\phi}^{\,({\rm int})}(\eta) \rg |^2$ for various states $|n \rg$, and of the expectation number of quanta excitations $< \hat{{\sf N}} >(\eta)=\sum_n n |c_n(\eta)|^2$.
\end{enumerate}

\begin{figure}[!t]
\begin{tabular}{cc}
\includegraphics[scale=0.85]{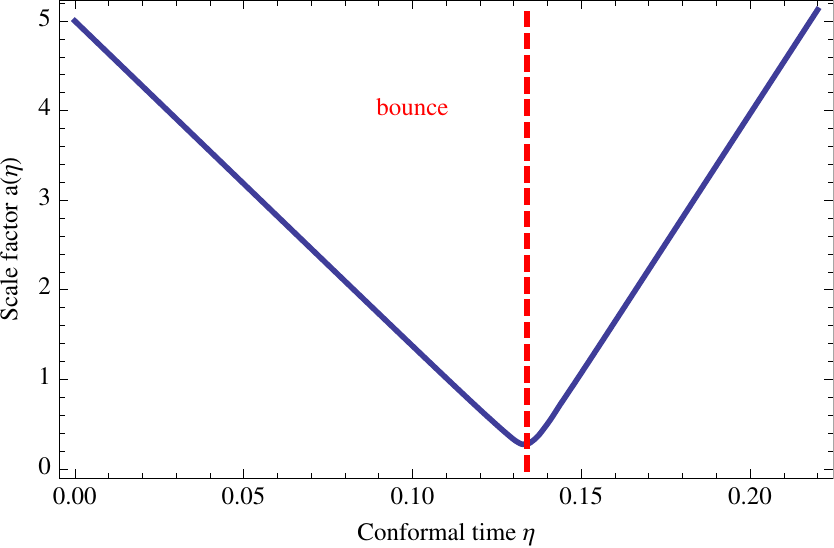} &
\includegraphics[scale=0.9]{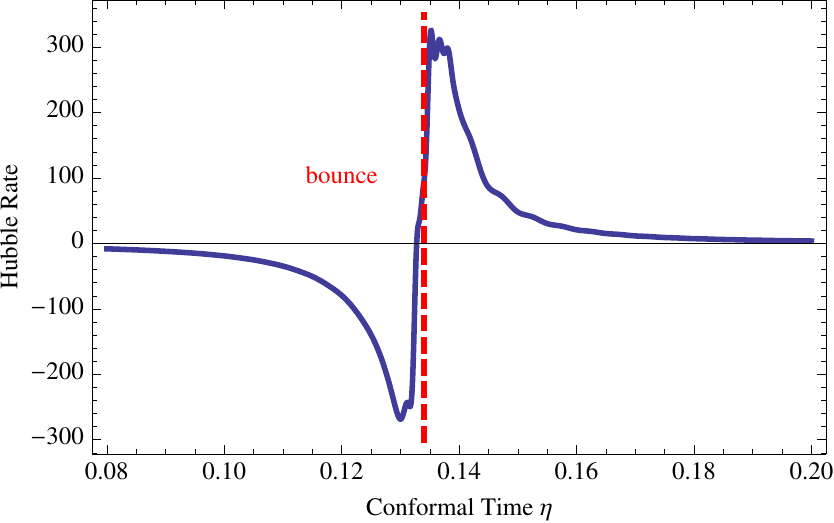}
\end{tabular}
\caption{The evolution of the scale factor $a(\eta)$ (left panel) and the Hubble rate (right panel) as a function of the conformal time $\eta$. The initial value of $a$ is $a_0=5$ and the initial state is $|\tilde{\phi}^{\,({\rm int})}_0 \rangle=|0 \rg$.  Note the change of slope, which increases after the bounce due to the production of the anisotropic energy and in accordance with the anticipation in \cite{B9vibronica}.} \label{figure1}
\end{figure}
\begin{figure}[!t]
\begin{tabular}{cc}
\includegraphics[scale=0.9]{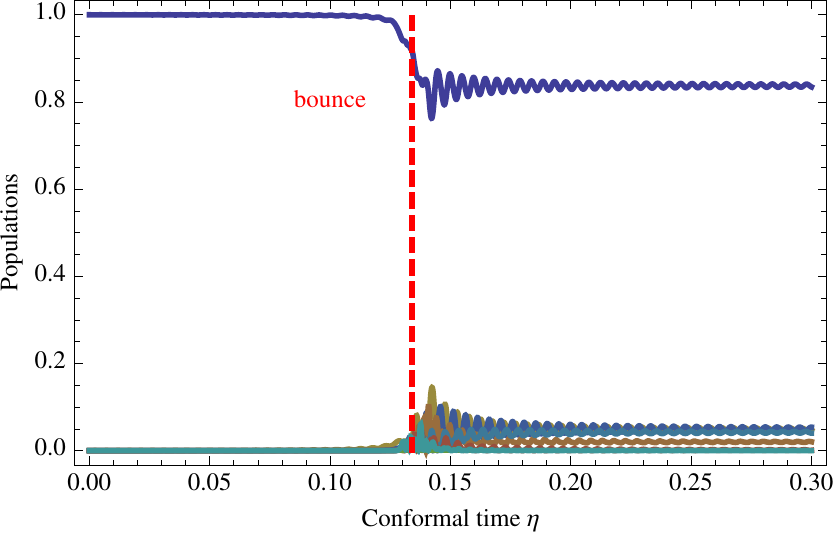} &
\includegraphics[scale=0.9]{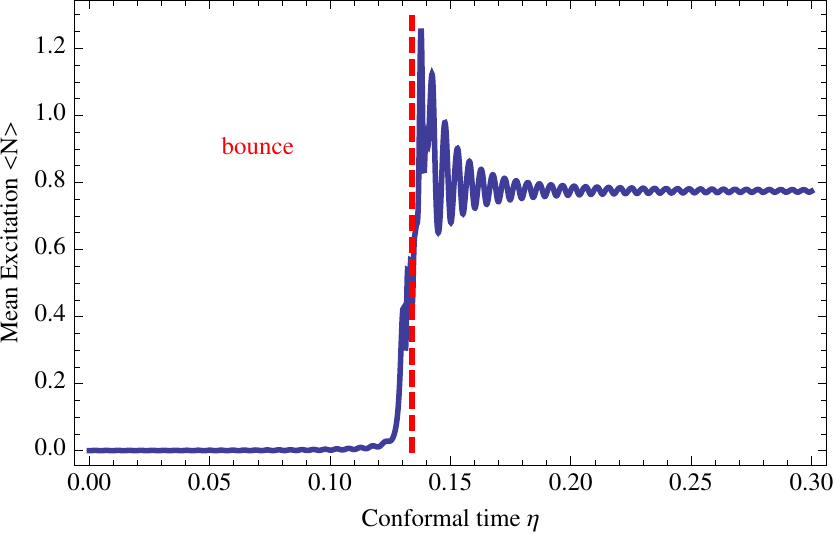}
\end{tabular}
\caption{Evolution of the quantum state with conformal time when the initial value of $a$ is $a_0=5$ and the initial state is $|\tilde{\phi}^{\,({\rm int})}_0 \rangle=|0 \rg$. On the left panel the evolution of the populations $|c_n(\eta)|^2$ for $n=0,1,\dots,12$. $|c_0(\eta)|^2$ corresponds to the curve on the top. On the right panel, the mean excitation $<\hat{{\sf N}}>(\eta)$.} \label{figure2}
\end{figure}

\subsection{Results}
\label{ressim}
\subsubsection{Stability of the ground state $| n=0 \rg$ during the bounce}
We begin with the study of the ground state stability by imposing $|\tilde{\phi}^{\,({\rm int})}_0 \rangle=|n=0 \rangle$. The numerical integration includes the first 12 eigenstates $|n=0 \rg$, $|n=1\rg$, $\dots$, $|n=12 \rg$. As expected from our previous paper \cite{B9vibronica}, we find that the vacuum remains essentially unchanged throughout the bounce. In other words, the evolution of the system is adiabatic in this case. This proves that the Friedmann model, quantum corrected by the zero-point energy of anisotropy degrees of freedom, is stable. This is the first main point of our results.

The figure \ref{figure1} shows the evolution of the scale factor, which exhibits the bounce. The figure \ref{figure2} proves the stability of the ground state. We notice that only some minor transitions between $|c_n(\eta)|^2$'s occur very close to the bounce.

\begin{figure}[!t]
\begin{tabular}{cc}
\includegraphics[scale=0.85]{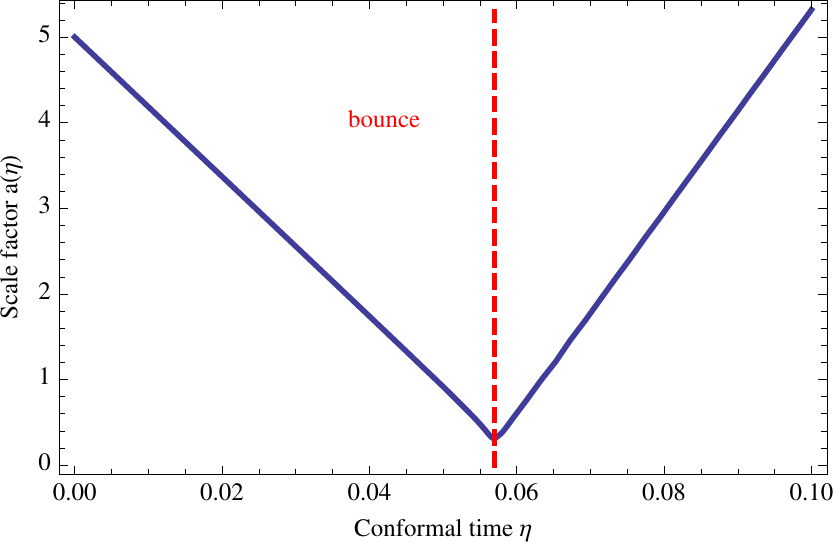} &
\includegraphics[scale=0.9]{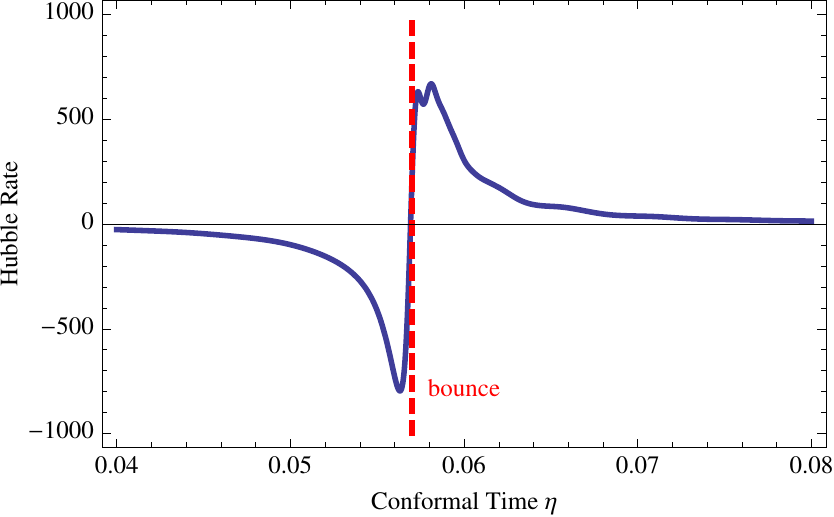}
\end{tabular}
\caption{The evolution of the scale factor $a(\eta)$ (left panel) and the Hubble rate (right panel) as a function of the conformal time $\eta$. The initial value of $a$ is $a_0=5$ and the initial anisotropy state is $|\tilde{\phi}^{\,({\rm int})}_0 \rangle=|n=2 \rg$.} \label{figure3}
\end{figure}
\begin{figure}[!t]
\begin{tabular}{cc}
\includegraphics[scale=0.85]{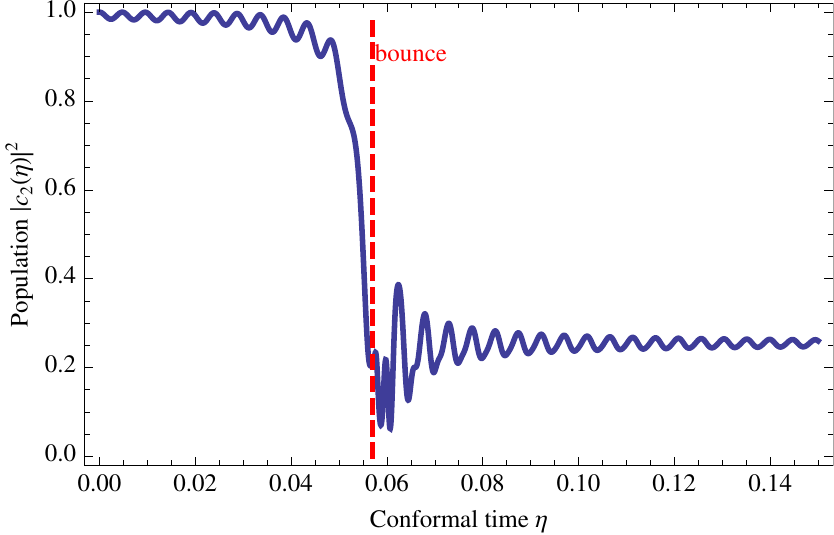} &
\includegraphics[scale=0.85]{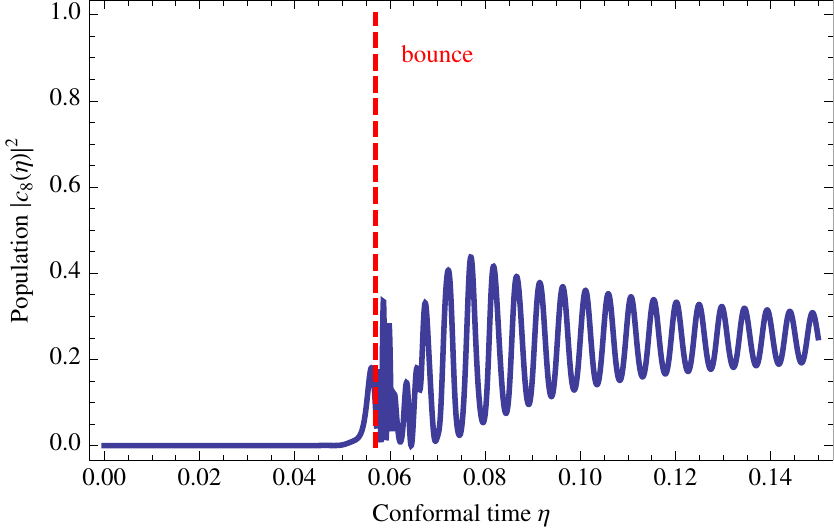} \\
\includegraphics[scale=0.85]{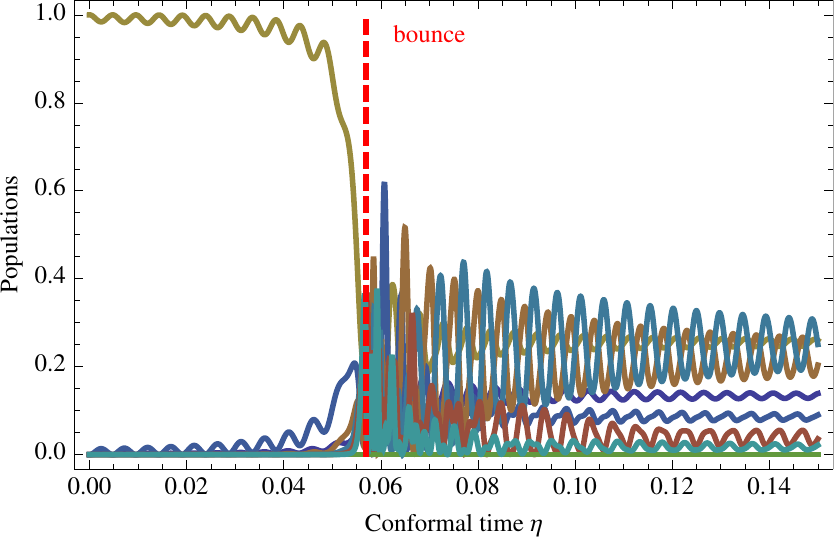} &
\includegraphics[scale=0.85]{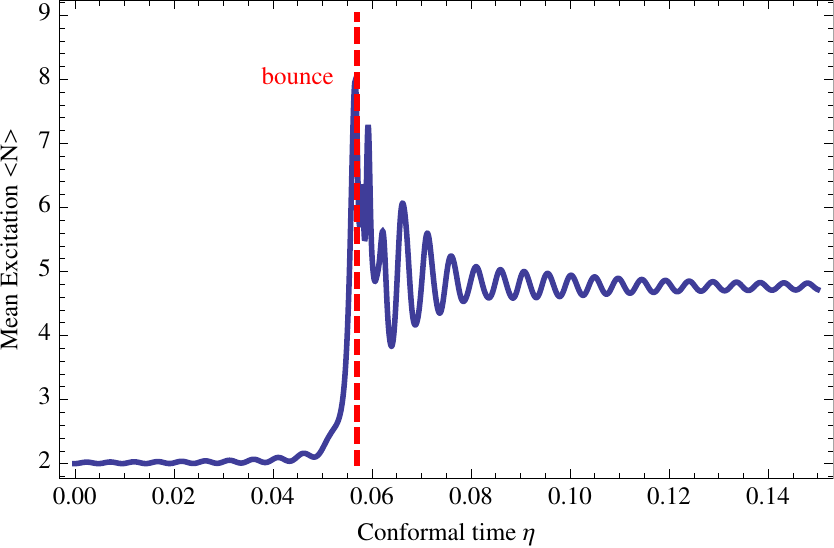}
\end{tabular}
\caption{Evolution of the quantum state in the conformal time for the initial value of $a$ is $a_0=5$ and the initial anisotropy state is $|\tilde{\phi}^{\,({\rm int})}_0 \rangle=|n=2 \rg$. In the top left panel the decay of the initial energy level $n=2$. In the top right panel the excitation of the energy level $n=8$. In the bottom left panel the evolution of the populations $|c_n(\eta)|^2$ for $n=0,1,\dots,12$. In the bottom right panel, the mean excitation $<\hat{{\sf N}}>(\eta)$.} \label{figure4}
\end{figure}

\subsubsection{Relative stability of low excited states during the bounce}
We address the stability of excited states by imposing $|\tilde{\phi}^{\,({\rm int})}_0 \rangle=|n=2 \rangle$ as the initial state. The numerical integration includes again the first 12 eigenstates $|n=0 \rg$, $|n=1\rg$, $\dots$, $|n=12 \rg$. We find that the excitations are spread by the bounce among the first 8 energy levels. In other words, the evolution of the system is not strictly adiabatic in this case, however the spreading does not involve a very large number of states.

The figure \ref{figure3} shows the evolution of the scale factor undergoing the bounce. We notice that the slopes in the plot are slightly different before and after the bounce: this is the effect of the backreaction of the anisotropic variables on the scale factor.

The figure \ref{figure4} illustrates the nonadiabatic behavior with the decay of the level $n=2$ and the excitation of other energy levels as $n=8$ for instance. The last panel shows that the mean expectation value $< \hat{{\sf N}} >$ reaches $< \hat{{\sf N}} > \simeq 5$ in the limit far from the bounce.

\begin{figure}[!ht]
\includegraphics[scale=0.9]{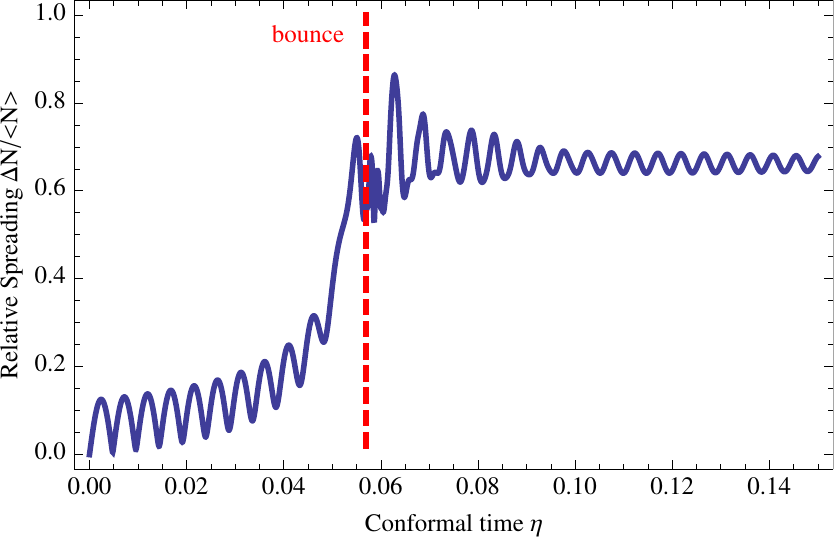}
\caption{Evolution of the relative spreading $\Delta N/<N>$ as a function of the conformal time $\eta$. The initial value of $a$ is $a_0=5$ and the initial state is $|\tilde{\phi}^{\,({\rm int})}_0 \rangle=|n=2 \rg$.} \label{figure5}
\end{figure}

The figure  \ref{figure5} illustrates the spreading of the state during its evolution in terms of $\Delta N/<N>$. We remark that the ratio reaches the asymptotic limit value $\Delta N \simeq 0.6 <N>$. Therefore, it contradicts $\Delta N \ll \, <N>$, the true adiabatic behavior, though the spreading is limited.

\section{\bf Conclusions}
\label{conclu}
In the present paper we have derived the complete and consistent set of equations of motion for the quantum mixmaster universe within the vibronic framework. It includes the entanglement between the isotropic and anisotropic components of the Hilbert space, resulting in backreation of the latter on the former. This framework is a natural extension of the adiabatic approximations studied by us before. The anisotropy variables are given a full quantum treatment, whereas the treatment of the isotropic variables is confined to semiclassical states in order to reproduce the classical theory away from the bounce. The Bianchi-IX anisotropy potential was approximated by the harmonic potential, which led to results quantitatively valid only in the deep quantum regime. Nevertheless, it is expected that the qualitative features of the approximated model, like singularity resolution, stability of the BO-type solutions, excitation of anisotropy in the remaining solutions and the subsequent phase of accelerated expansion, correspond to the behavior of the exact model.

This approach comprises, as one of the equations of motion, the modified Friedmann equation, similar to the one obtained previously for the adiabatic solutions \cite{Bergeron2015long}. Nevertheless, the coefficients of the present Friedmann-like equation are time-dependent and their evolution is governed by the supplementary Schr\"odinger equation. We resorted to numerical simulations in order to investigate the singularity resolution and stability issues in this extended framework.
    
We studied numerically the stability of the ground state, which corresponds to the quantum Friedmann universe, and the first few excited states. In agreement with the adiabatic approximations \cite{Bergeron2015short,Bergeron2015long}, we found that the ground state of our model remains stable throughout the bounce even if mixing between isotropic and anisotropic states is allowed. The evolution of the slightly excited eigenstates is no longer adiabatic and the stability of the bounce is apparently weakened. We observe the effect of the backreaction of anisotropic variables on the isotropic ones through the slope change in the plot of the semiclassical averaged scale factor in the conformal time, which takes place just after the bounce due to the anisotropy production.
    
Next steps could include taking into account higher excited states to see if the breaking of stability gets even more robust and manifest. However, increasing the excitation number also makes the quantitative predictions of our model less reliable because of the employed harmonic approximation. Ultimately, we will need to include the exact anisotropy potential of the mixmaster universe, or a more accurate approximation, which is possibly achievable only with numerical solutions to the respective eigenvalue problem.
    
Concerning the possible extension of the presented framework, the bouncing scenario might be considered as a scattering process. As in the far past and future, for large values of the scale factor $a$, the anisotropy potential vanishes, it is possible to reformulate the whole evolution in the interaction picture, which prompts a suitable definition of the $S$-matrix. Such an approach should provide enough information on the origin of the present classical space-time, viewed as an asymptotic state away from the regime of strong quantum interaction, into which another classical and formerly contracting space-time had collapsed.

\section*{Acknowledgements}
P.M. was supported by the ``Mobilno\'s\'c Plus" fellowship from Ministerstwo Nauki i Szkolnictwa Wy\.zszego.

\end{document}